\documentclass[conference]{IEEEtran}
\usepackage{cite}

\ifCLASSINFOpdf
\else
\fi
  \usepackage[caption=false,font=footnotesize]{subfig}
\usepackage{amsmath,amssymb,amsfonts,amsthm}
\usepackage{algorithmic}
\usepackage{graphicx}
\usepackage{textcomp}
\usepackage{xcolor}
\usepackage{abbrevs}
\usepackage[normalem]{ulem}

\newtheorem{theorem}{Theorem}

\newcommand{\dif}{\mathrm{d}}

\newcommand{\traj}{X} 
\newcommand{\stseq}{x} 
\newcommand{\tb}{\beta} 
\newcommand{\td}{\varepsilon} 
\newcommand{\tlen}{\ell} 
\newcommand{\targetStateSpace}{\mathcal{X}} 

\newcommand{\timeseq}[2]{#1:#2}
\newcommand{\timeset}[2]{\mathbb{N}_{#1}^{#2}}
\newcommand{\trajStateSpace}[1]{\mathcal{T}_{#1}} 
\newcommand{\existencespace}[1]{I_{#1}}

\newcommand{\measStateSpace}{\mathcal{Z}} 
\newcommand{\trackTable}{\mathbb{T}}

\newcommand{\thepapertitle}{Spatiotemporal Constraints for Sets of Trajectories with Applications to PMBM Densities}

\allowdisplaybreaks

\begin{document}
%
\title{\thepapertitle}
%
%
%

\author{\IEEEauthorblockN{Karl Granstr\"om, Lennart Svensson\\ Yuxuan Xia}
\IEEEauthorblockA{\textit{Dept. of Electrical Eng.} \\
\textit{Chalmers Univ. of Tech.}\\
Gothenburg, Sweden \\
{\footnotesize\texttt{firstname.lastname@chalmers.se}}}
\and
\IEEEauthorblockN{\'Angel F. Garc\'ia-Fern\'andez}
\IEEEauthorblockA{\textit{Dept. of Electrical Eng. and Electronics} \\
\textit{Univ. of Liverpool}\\
Liverpool, UK \\
{\footnotesize\texttt{angel.garcia-fernandez@liverpool.ac.uk}}}
\and
\IEEEauthorblockN{Jason Williams}
\IEEEauthorblockA{\textit{CSIRO} \\
\textit{Data 61}\\
Brisbane, QLD, Australia \\
{\footnotesize\texttt{jason.williams@data61.csiro.au}}}
}

\maketitle

\begin{abstract}
In this paper we introduce spatiotemporal constraints for trajectories, i.e., restrictions that the trajectory must be in some part of the state space (spatial constraint) at some point in time (temporal constraint). Spatiotemporal contraints on trajectories can be used to answer a range of important questions, including, e.g., \emph{``where did the person that were in area A at time t, go afterwards?''}. We discuss how multiple constraints can be combined into sets of constraints, and we then apply sets of constraints to set of trajectories densities, specifically Poisson Multi-Bernoulli Mixture (\pmbm) densities. For Poisson target birth, the exact posterior density is \pmbm for both point targets and extended targets. In the paper we show that if the unconstrained set of trajectories density is \pmbm, then the constrained density is also \pmbm. Examples of constrained trajectory densities motivate and illustrate the key results.
\end{abstract}


%
\IEEEpeerreviewmaketitle

\section{Introduction}

Multiple Target Tracking (\mtt) can be defined as the processing of noisy sensor measurements to determine 1) the number of targets, and 2) each target's trajectory, see, e.g., \cite{BarShalomWT:2011}. In this paper we consider the standard \mtt models, see, e.g., \cite[Sec. 10.2]{Mah07}, where the birth is Poisson. When the birth model is Poisson, the Poisson Multi-Bernoulli Mixture (\pmbm) density is multi-object conjugate for both point target measurements \cite{Wil12} and for extended target measurements \cite{GranstromFS:2020}. The resulting \pmbm filters have a structure with multiple hypotheses corresponding to different data associations, however, track continuity was not established.

In previous work, track continuity has been established using labels, see, e.g., \cite{GarciaFernandezGM:2013,AokiMSBB:2016,VoVo13}, or related approaches, see, e.g., \cite{PantaCV2009,HoussineauC:2018}. With labels, trajectories are formed by connecting estimates from different times that have the same label. The \dglmb filter \cite{VoVo13} is conjugate for labelled multi-Bernoulli birth; the similarities and differences between the \pmbm and \dglmb conjugate priors are discussed in, e.g., \cite[Sec. V.C]{GranstromFS:2020} and \cite[Sec. IV]{GarciaFernandezWGS:2018}.

Several simulation studies have shown that, compared to tracking filters built upon labelled \rfs, the \pmbm filters provide state-of-the-art performance for tracking the set of targets, see, e.g., \cite{XiaGSGF:2017,GarciaFernandezWGS:2018,XiaGSGF:2018,GranstromFS:2016fusion,GranstromFS:2020,XiaGSGFW:2019,XiaGSGFW:jaifMultiScanPMBMtrackers}. \pmbm filters are versatile, and have been used with data from lidars, radars and cameras \cite{GranstromRFS:2017,CamentACP:2017,CamentAC:2018,GranstromSRXF:2018,ScheideggerG:2018,MotroG:2018}, and are applicable to both tracking and mapping \cite{FatemiGSRH:2016_PMBradarmapping}, as well as joint tracking and sensor localisation \cite{FrohleLGW:multisensorPMB}

In this paper we rely on modelling the \mtt problem using random finite sets (\rfs) of trajectories \cite{SvenssonM:2014,GarciaFernandezSM:2019}. Within this set of trajectories framework, the goal of Bayesian \mtt is to compute the posterior density over the set of trajectories. For both the standard point target model and the standard extended target model, it has been shown that the set of trajectories \pmbm density is multi-object conjugate \cite{GranstromSXGFW:2018,GranstromSXGFW:PMBMtrackersARXIV,XiaGSGFW:2019}, and it has been shown that regardless of what time intervals we consider for the trajectories and the measurements, the set of trajectories density is \pmbm \cite{GranstromSXGFW:PMBMtrackersARXIV}. 

\begin{figure*}
	\centering

	\subfloat[]{\label{fig:all_trajectories}\includegraphics[width=0.66\columnwidth]{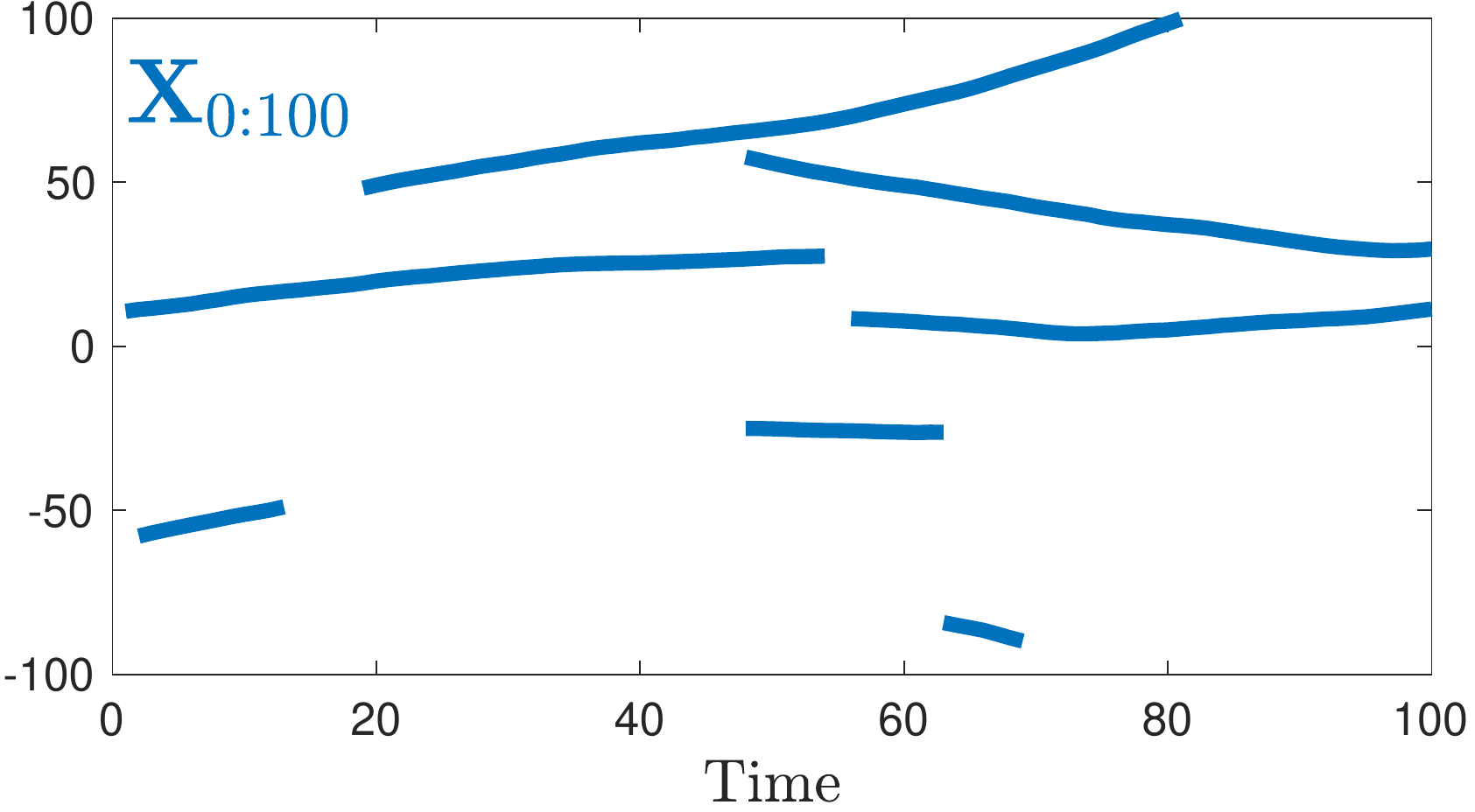}}\hfil\subfloat[]{\label{fig:consecutive_time_constraints}\includegraphics[width=0.66\columnwidth]{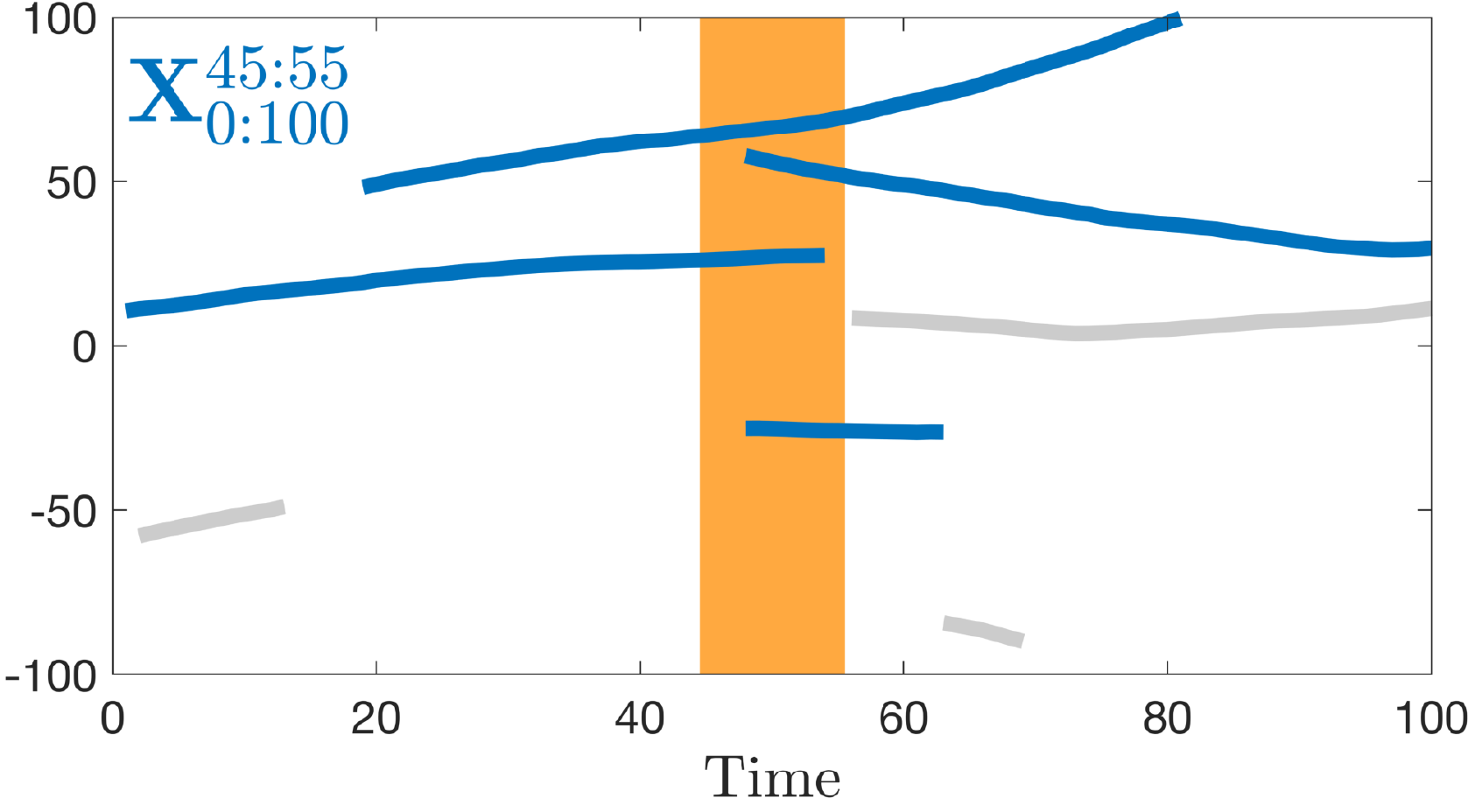}}\hfil\subfloat[]{\label{fig:single_constraint}\includegraphics[width=0.66\columnwidth]{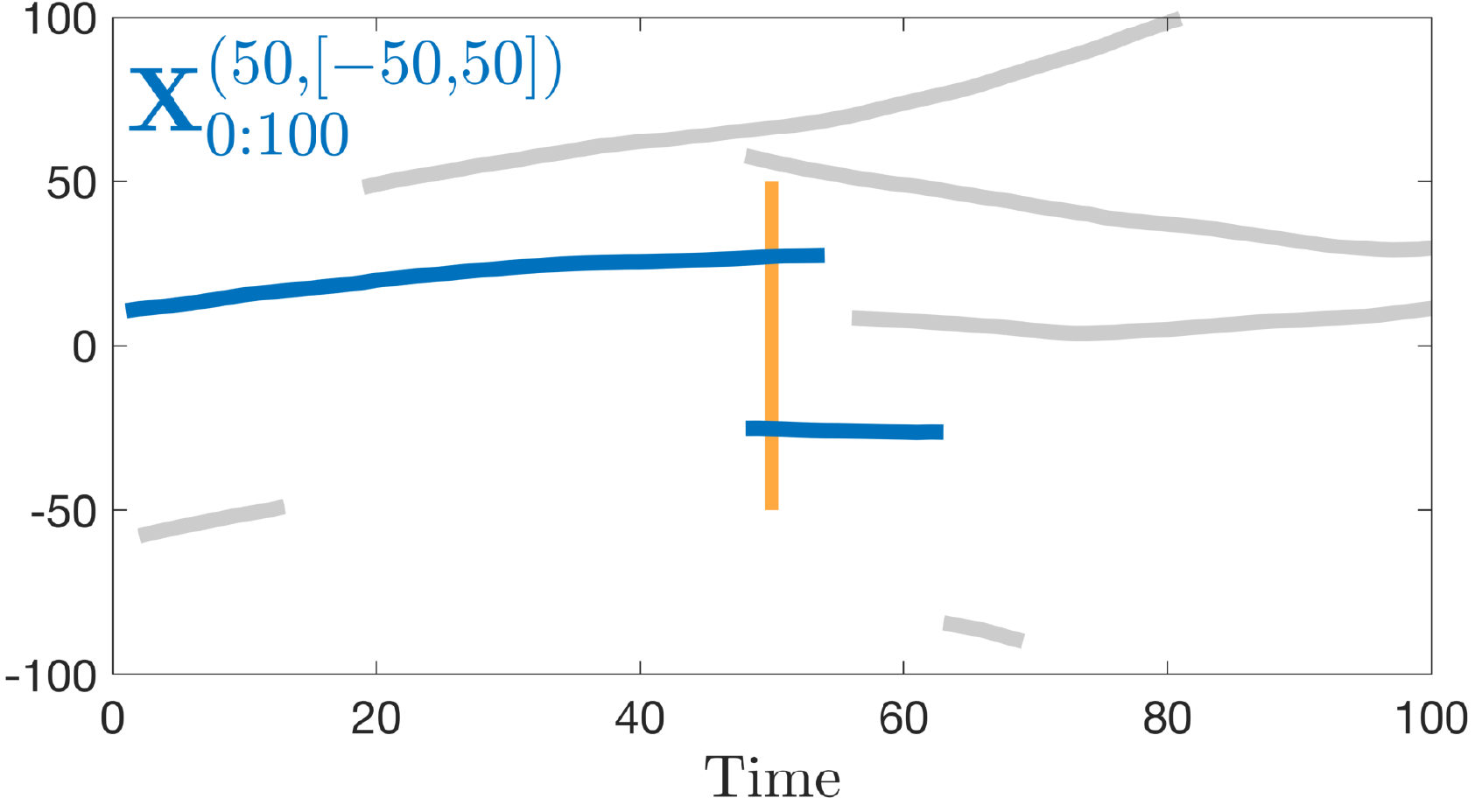}}

	\caption{(a) Seven trajectories in the time window $\timeseq{0}{100}$, in a 1D state space with surveillance area $[-100,\ 100]$. (b) Example of a constraint in the form of a window in time. Four of the trajectories (blue) are alive for at least one time step in the time window (orange). Remaining three of the trajectories (gray) are not alive at any time step in the time window, i.e., the do not meet the time window constraint, and are therefore not included in the time window constrained set of trajectories $\settraj_{\timeseq{0}{100}}^{\timeseq{45}{55}}$. (c) Example of a single spatiotemporal constraint applied to the set of trajectories from Figure~\ref{fig:all_trajectories}. Constraint $\trajconstraint = (50, [-50, \ 50])$ illustrated in orange, trajectories that meet the constraint illustrated in blue, trajectories that do not meet the constraint illustrated in grey.}
	
\end{figure*}

In this paper we introduce spatiotemporal constraints for set of trajectories, where a spatiotemporal constraint is a constraint in both the target state space (spatial) and in time (temporal). In Figure~\ref{fig:all_trajectories} an example with seven trajectories in the time interval $\timeseq{\alpha}{\gamma}=\timeseq{0}{100}$ are shown; here the state space is $\targetStateSpace = \mathbb{R}$ and the field of view is $[-100, \ 100]$. By applying spatiotemporal constraints to sets of trajectories, and set of trajectories densities, we can provide answers to questions such as:
\begin{itemize}
	\item What is the distribution of the trajectories that passed through an area of interest sometime during time interval of interest?
	\item In a certain time interval, what is the distribution of the trajectories that lingered in the area of interest for at least some minimum number of time steps?
	\item What is the distribution of the trajectories that were in area A in the first time interval, and in area B in the second interval?
\end{itemize}

In addition to defining what a spatiotemporal trajectory constraint is, we discuss how multiple constraints can be combined into sets of constraints, and we show that if we have a \pmbm set of trajectories density, then the constrained density is also \pmbm, i.e., application of sets of spatiotemporal trajectory constrains maintains the functional form of the density. This is important, because the conjugacy of the \pmbm density is key to enabling the development of tracking and mapping algorithms that are both accurate and computationally efficient. 

The paper is organised as follows. In the next section we provide some background about sets of trajectories. Spatiotemporal trajectory constraints are defined and discussed in Section~\ref{sec:spatiotemporal_trajectory_constraints}. Constrained \pmbm densities are presented in Section~\ref{sec:spatiotemporally_constrained_PMBM_density}. Some examples are presented in Section~\ref{sec:examples}, and the paper is concluded in Section~\ref{sec:conclusion}.

\section{Sets of trajectories background}

Let $x_{k}\in\targetStateSpace$ denote a target state at discrete time $k$, where $\targetStateSpace$ represent the base state space, and let $z_{k}\in\measStateSpace$ denote a measurement at time $k$, where $\measStateSpace$ is the measurement state space. 
We utilise the standard point object models. The dynamics model has Poisson Point Process (\ppp) birth with intensity $\lambda^\mathrm{b}(x_k)$, probability of survival $P^{\rm S}(x_k)$, and single-target transition density $\pi^{x}(x_k|x_{k-1})$. The measurement model has probability of detection $P^{\rm D}(x_k)$, each measurement is from at most one target with measurement likelihood $\varphi^{z}(z_k|x_k)$, and clutter is \ppp with intensity $\lambda^\mathrm{FA}(z_k)$.

For two time steps $\alpha$ and $\gamma$, $\alpha\leq\gamma$, following standard tracking notation the ordered sequence of consecutive time steps is denoted $\timeseq{\alpha}{\gamma} = \left( \alpha,\alpha+1,\ldots,\gamma-1,\gamma \right)$. The (unordered) set of consecutive time steps is denoted $\timeset{\alpha}{\gamma} = \left\{\alpha,\alpha+1,\ldots,\gamma-1,\gamma\right\}$. We use the trajectory state model proposed in \cite{SvenssonM:2014,GarciaFernandezSM:2019}, in which the trajectory state is a tuple
\begin{align}
	\traj = \left(\tb,\td,\stseq_{\timeseq{\tb}{\td}}\right)
\end{align}
where $\tb$ is the discrete time step of the trajectory birth, i.e., the time step when the trajectory begins; $\td$ is the discrete time step of the trajectory's most recent state, i.e., the time step when the trajectory ends; and $\stseq_{\timeseq{\tb}{\td}}$ is, given $\tb$ and $\td$, the sequence of states 
\begin{subequations}
	\begin{align}
		& \left(x_{\tb},x_{\tb+1},\ldots,x_{\td-1},x_{\td}\right), \\
		& x_{k}\in\targetStateSpace,\ \forall k\in\timeset{\tb}{\td}.
	\end{align}
\end{subequations}
The length of a trajectory $\traj$ is $\tlen = \td-\tb+1$ time steps; $\tlen$ is finite because $\tb$ and $\td$ are finite. 

For two time steps $\alpha$ and $\gamma$, $\alpha\leq\gamma$, the trajectory state space for trajectories in the time interval $\timeseq{\alpha}{\gamma}$ is \cite{GranstromSXGFW:PMBMtrackersARXIV}
\begin{align}
	& \trajStateSpace{\timeseq{\alpha}{\gamma}} = \uplus_{(\tb,\td)\in \existencespace{\timeseq{\alpha}{\gamma}}} \{\tb\}\times\{\td\}\times\targetStateSpace^{\td-\tb+1},
\end{align}
where $\uplus$ denotes union of disjoint sets, $\existencespace{\timeseq{\alpha}{\gamma}} = \{ (\tb,\td) : \alpha\leq \tb \leq \td \leq \gamma \}$ and $\targetStateSpace^{\tlen}$ denotes $\tlen$ Cartesian products of $\targetStateSpace$. The finite lengths of trajectories in $\trajStateSpace{\timeseq{\alpha}{\gamma}}$ are restricted to $1\leq\tlen\leq\gamma-\alpha+1$.

The trajectory state density factorises as follows
\begin{align}
	p(\traj) = p(\stseq_{\timeseq{\tb}{\td}} | \tb,\td) P(\tb,\td), \label{eq:trajectory_state_density}
\end{align}
where the domain of $P(\tb,\td)$ is $\existencespace{\timeseq{\alpha}{\gamma}}$ for $\traj \in \trajStateSpace{\timeseq{\alpha}{\gamma}}$. Integration is performed as follows \cite{GarciaFernandezSM:2019},

\begin{align}
	& \int_{\trajStateSpace{\timeseq{\alpha}{\gamma}}} p(\traj) \diff \traj \label{eq:trajectoryDensityIntegration}\\
	& = \sum_{(\tb,\td)\in\existencespace{\timeseq{\alpha}{\gamma}}} \left[ \int_{\targetStateSpace^{\tlen}} p(\stseq_{\timeseq{\tb}{\td}} | \tb,\td) \diff \stseq_{\timeseq{\tb}{\td}} \right] P(\tb,\td) . \nonumber
\end{align}%
For any subset of time steps $\mathbb{M}_{\tb}^{\td} \subseteq \mathbb{N}_{\tb}^{\td}$, we denote by $p(\stseq_{\mathbb{M}_{\tb}^{\td}}|\tb,\td)$ the conditional density of the states for times in $\mathbb{M}_{\tb}^{\td}$, i.e., all other time steps in $ \mathbb{N}_{\tb}^{\td}$ have been marginalized from the density.


The set of trajectories in the time interval ${\timeseq{\alpha}{\gamma}}$ is denoted as $\settraj_{\timeseq{\alpha}{\gamma}}$. The domain for $\settraj_{\timeseq{\alpha}{\gamma}}$ is $\mathcal{F}(\trajStateSpace{\timeseq{\alpha}{\gamma}})$, the set of all finite subsets of $\trajStateSpace{\timeseq{\alpha}{\gamma}}$. In some applications we consider a subset of the trajectories in $\settraj_{\timeseq{\alpha}{\gamma}}$, namely the ones that were alive at some point in the time interval $\timeseq{\eta}{\zeta}$, where $\alpha\leq\eta\leq\zeta\leq\gamma$. We denote this set of trajectories as
\begin{align}
	\settraj_{\timeseq{\alpha}{\gamma}}^{\timeseq{\eta}{\zeta}} = \left\{ \traj = \left(\tb,\td,\stseq_{\timeseq{\tb}{\td}}\right) \in\trajStateSpace{\timeseq{\alpha}{\gamma}} \ : \ \timeset{\tb}{\td}\cap \timeset{\eta}{\zeta} \neq\emptyset \right\}. \label{eq:temporallyConstrainedSetOfTrajectories}
\end{align}

The multi-trajectory density $f\left(\settraj_{\timeseq{\alpha}{\gamma}}\right)$ is defined analogously to the multi-target density. Let $g\left(\settraj_{\timeseq{\alpha}{\gamma}}\right)$ be a real-valued function on a set of trajectories $\settraj_{\timeseq{\alpha}{\gamma}}$. Integration over sets of trajectories is defined as regular set integration \cite{Mah07}:
\begin{align}
	& \int{g\left(\settraj_{\timeseq{\alpha}{\gamma}}\right) \delta \settraj_{\timeseq{\alpha}{\gamma}}} \triangleq g(\emptyset) + \nonumber\\ 
	& \quad \sum_{n=1}^{\infty}\frac{1}{n!}\idotsint{g(\{\traj^1,\dots,\traj^n\})\dif \traj^1 \cdots \dif \traj^n} . \label{eq:SetTrajIntegral}
\end{align}
The multi-target Dirac delta is defined as \cite[Sec. 11.3.4.3]{Mah07}
\begin{align}
	\delta_{\setX'}(\setX) \triangleq 
		\begin{cases}
			0 & |\setX| \neq |\setX'| \\
			1 & \setX = \setX' = \emptyset \\
			\displaystyle\sum_{\sigma\in\Gamma_{n}} \prod_{i=1}^{n} \delta_{\traj_{\sigma_{i}}'}(\traj_i) & 
			\begin{cases}
				\setX = \left\{\traj_{i}\right\}_{i=1}^{n} \\
				\setX' =  \left\{\traj_{i}'\right\}_{i=1}^{n}
			\end{cases}
		\end{cases}
\end{align}
Here, the trajectory Dirac delta is defined as
\begin{align}
	\delta_{\traj'}(\traj) = \Delta_{\tb'}(\tb) \Delta_{\td'}(\td) \delta_{\stseq_{\timeseq{\tb'}{\td'}}'}(\stseq_{\timeseq{\tb}{\td}})
\end{align}
where $\Delta_{y'}(y)$ is Kronecker delta, defined as
\begin{align}
	\Delta_{y'}(y) = \begin{cases} 1 & y=y' \\ 0 & \text{otherwise} \end{cases}
\end{align}
and $\delta_{y'}(y)$ is Dirac delta for continuous variables, defined by the property
\begin{align}
	\int p(y) \delta_{y'}(y) \diff y = p(y').
\end{align}
For a sub-set of the base state space $\statespaceconstraint \subseteq \targetStateSpace$, $\statespaceconstraint^{\complement}$ is its complement, such that $\statespaceconstraint \cup \statespaceconstraint^{\complement} = \targetStateSpace$.

%
%

\section{Spatiotemporal trajectory constraints}
\label{sec:spatiotemporal_trajectory_constraints}

In this section we give a general description of the problem considered in this paper: spatiotemporal constraints applied to sets of trajectories and densities on sets of trajectories, especially the \pmbm density. Constraints in the form of windows in time were introduced in \cite{GranstromSXGFW:PMBMtrackersARXIV}, where the set of trajectories in the time interval $\timeseq{\alpha}{\gamma}$ were restricted to the trajectories alive at some point in the time interval $\timeseq{\eta}{\zeta}$, cf. \eqref{eq:temporallyConstrainedSetOfTrajectories}. An example restriction is shown in Figure~\ref{fig:consecutive_time_constraints}.

The problem considered in this paper is to generalise the time-window constraint to sets of spatiotemporal constraints, where each constraint specifies a time step (temporal constraint) and an area of the state space (spatial constraint), and where the set of constraints are not restricted to be consecutive in time, i.e., the temporal constraints are not required to form a window in time.

In the following subsection, we begin by giving a definition of a spatiotemporal trajectory constraint, and then proceed to discuss how a set of constraints, i.e., multiple constraints, can be combined.

\subsection{Trajectory constraint}
\label{sec:TrajectoryConstraint}
A spatiotemporal trajectory constraint $\trajconstraint$ consists of a time step $\timeconstraint$ (temporal constraint) and a region of the state space $\statespaceconstraint\subseteq\targetStateSpace$ (spatial constraint),
\begin{align}
	\trajconstraint = (\timeconstraint, \ \statespaceconstraint) . \label{eq:trajectoryConstraint}
\end{align}
For a time interval $\timeseq{\alpha}{\gamma}$ and a constraint $\trajconstraint$ where $\timeconstraint\in\timeset{\alpha}{\gamma}$, the set of trajectories in the time interval that fulfil the constraint is denoted $\settraj_{\timeseq{\alpha}{\gamma}}^{\trajconstraint}$ and is defined as
\begin{align}
	\settraj_{\timeseq{\alpha}{\gamma}}^{\trajconstraint} = \left\{ \traj\in\trajStateSpace{\timeseq{\alpha}{\gamma}} : \timeconstraint \in \timeset{\tb}{\td} \text{ and } x_{\timeconstraint} \in \statespaceconstraint \right\}.
	\label{eq:singleConstraintSet}
\end{align}
In other words, the trajectory constraint $\trajconstraint$ means that the trajectory must be alive at the time of the constraint ($\timeconstraint \in \timeset{\tb}{\td}$) and that the state at that time must be in the constraint region ($x_{\timeconstraint} \in \statespaceconstraint$).
In Figure~\ref{fig:single_constraint} we illustrate the set of trajectories in Figure~\ref{fig:all_trajectories} after applying the trajectory constraint $\trajconstraint = (50, [-50, \ 50])$.
%
%

\subsection{Sets of constraints}
\label{sec:SetOfTrajectoryConstraints}
Let us now consider a set of trajectory constraints \eqref{eq:trajectoryConstraint}, denoted $\setconstraint$. Note that, without loss of generality, we can assume that for any two constraints $\trajconstraint\in\setconstraint$ and $\trajconstraint'\in\setconstraint$ it holds that $\timeconstraint \neq \timeconstraint'$, because if $\timeconstraint = \timeconstraint'$, then the two constraints can be replaced by a single constraint $\trajconstraint'' = (\timeconstraint, \statespaceconstraint \cup \statespaceconstraint')$. For the sake of notational simplicity, we introduce an index set $\mathbb{I}$ for the constraints,
\begin{align}
	\setconstraint = \left\{ \trajconstraint_{i} \right\}_{i\in\mathbb{I}} = \left\{ (\timeconstraint_{i},\statespaceconstraint_{i}) \right\}_{i\in\mathbb{I}}.
\end{align}
It is important to note that the time constraints $\timeconstraint_{i}$ are not restricted to being consecutive, and the spatial constraints $\statespaceconstraint_{i}$ are not restricted to being overlapping, i.e., their pairwise intersections may be empty. The set of time step constraints is denoted $\settimeconstraint = \left\{ \timeconstraint_{i} \right\}_{i\in\mathbb{I}}$. Given a trajectory $\traj$ with time of birth $\tb$ and time of death $\td$, the index set for the constraints whose times $\timeconstraint$ fall inside the life-span of the trajectory, i.e., $\timeseq{\tb}{\td}$, and the corresponding set of time constraints, are denoted
\begin{align}
	\mathbb{I}_{\tb}^{\td} & = \left\{ i \in \mathbb{I} : \tb \leq \timeconstraint_{i} \leq \td \right\} , \\
	\settimeconstraint_{\tb}^{\td} & = \settimeconstraint \cap \timeset{\tb}{\td} = \left\{ \timeconstraint_{i} \right\}_{i\in\mathbb{I}_{\tb}^{\td}} .
\end{align}

With a single constraint, it is easy to determine which trajectories satisfy the constraint: it is the trajectories that are alive at the time of the constraint, and whose state is located in the constraint-region, see \eqref{eq:singleConstraintSet}. With a set of constraints, it is not necessarily trivial to answer the question: which trajectories satisfy the set of constraints? The reason for this is that there is a large number of ways in which the multiple constraints can be combined. We can enforce each trajectory constraint strictly, i.e., to satisfy the set of constraints, the trajectory has to satisfy every single constraint in the set of constraints. Alternatively, we can enforce any subset or combination of the constraints, e.g., to satisfy the set of constraints, a trajectory has to satisfy at least one of the constraints in the set. Which alternative is relevant is highly dependent on the type of application that is considered, and, arguably, there is no single correct answer.

It is not possible to consider all possible ways to combine multiple constraints in this paper, and we focus on two alternatives:
\begin{enumerate}
	\item Conjunct constraints: The trajectory satisfies all constraints in the time interval $\timeseq{\tb}{\td}$:
		\begin{align}
			\settimeconstraint_{\tb}^{\td} \neq \emptyset \text{, and } x_{\timeconstraint_{i}} \in \statespaceconstraint_{i}, \  \forall i \in \mathbb{I}_{\tb}^{\td} .
		\end{align}
		In other words, for the constraint times that the trajectory is alive, the trajectory is in the constrained state space regions for all of the time steps. We refer to this as a set of \emph{``conjunct''} constraints. The set of trajectories that fulfil a set of conjunct constraints is
		\begin{align}
			\settraj_{\timeseq{\alpha}{\gamma}}^{\setconstraint,c} = \left\{ \traj\in\trajStateSpace{\timeseq{\alpha}{\gamma}} \right. : \ & \settimeconstraint_{\tb}^{\td} \neq \emptyset \text{, and }  \\
			& \left. x_{\timeconstraint_{i}} \in \statespaceconstraint_{i}, \  \forall i \in \mathbb{I}_{\tb}^{\td} \right\}. \nonumber 
		\end{align}
		Conjunct constraints are relevant, e.g., when one is interested in objects that linger in an area of interest, i.e., the state has to satisfy the spatial constraint for all relevant time steps.
		
	\item Disjunct constraints: The trajectory satisfies at least one constraint:
		\begin{align}
			\settimeconstraint_{\tb}^{\td} \neq \emptyset \text{, and } \exists i \in \mathbb{I}_{\tb}^{\td} : x_{\timeconstraint_{i}} \in \statespaceconstraint_{i} .
		\end{align}
		In other words, for the constraint times that the trajectory is alive, the trajectory is in the constrained state space region for at least one of the time steps. We refer to this as a set of \emph{``disjunct''} constraints. The set of trajectories that fulfil a set of disjunct constraints is
		\begin{align}
			\settraj_{\timeseq{\alpha}{\gamma}}^{\setconstraint,d} = \left\{ \traj\in\trajStateSpace{\timeseq{\alpha}{\gamma}} \right. : \ & \settimeconstraint_{\tb}^{\td} \neq \emptyset \text{, and }  \\
			& \left. \exists i \in \mathbb{I}_{\tb}^{\td} : x_{\timeconstraint_{i}} \in \statespaceconstraint_{i} \right\}. \nonumber 
		\end{align}
		Disjunct constraints are relevant, e.g., when one is interested in objects that pass through an area of interest, i.e., the state only has to satisfy the spatial constraint for one of the relevant time steps.
\end{enumerate}%

\begin{figure*}
	\centering

	\includegraphics[width=0.66\columnwidth]{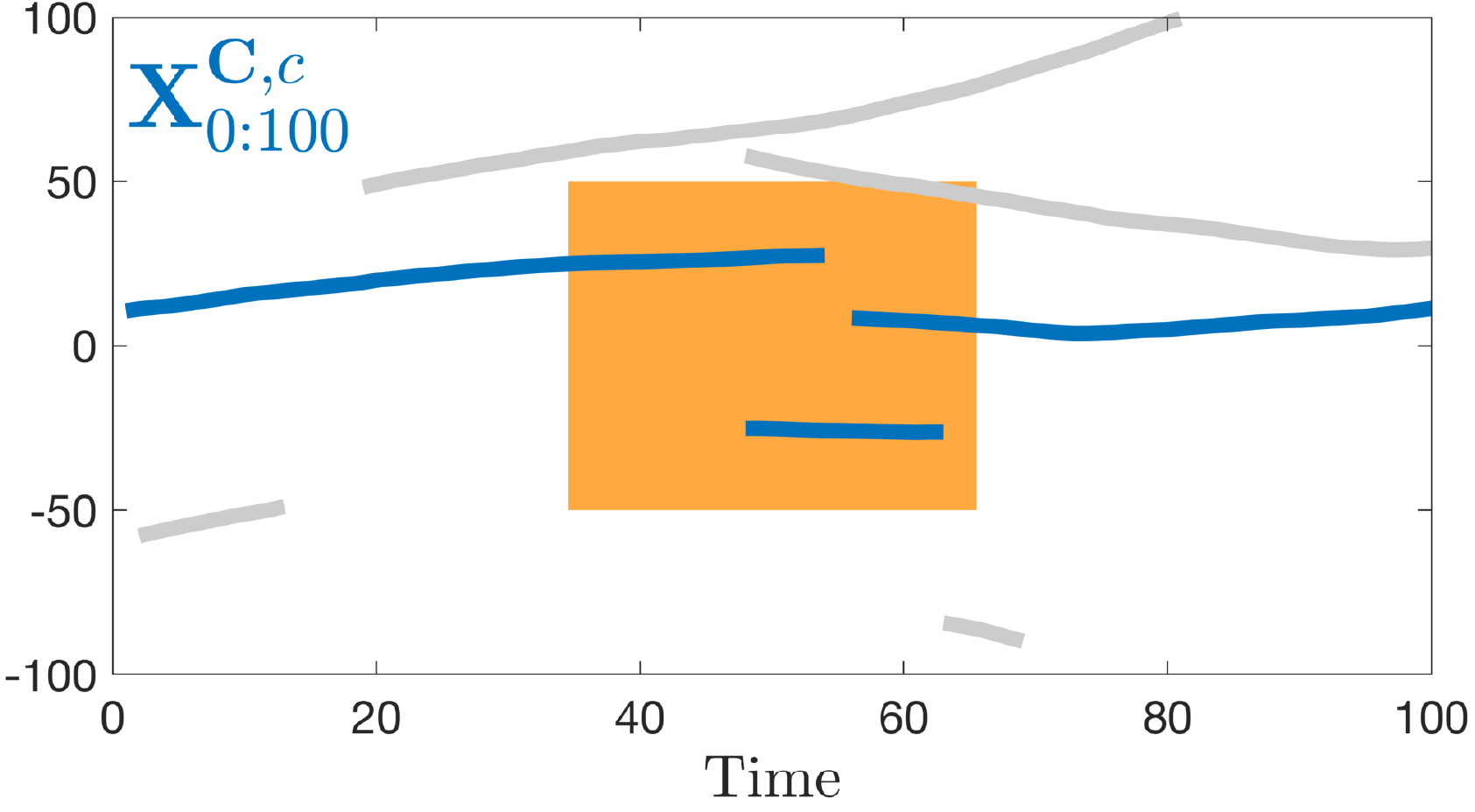}
	\includegraphics[width=0.66\columnwidth]{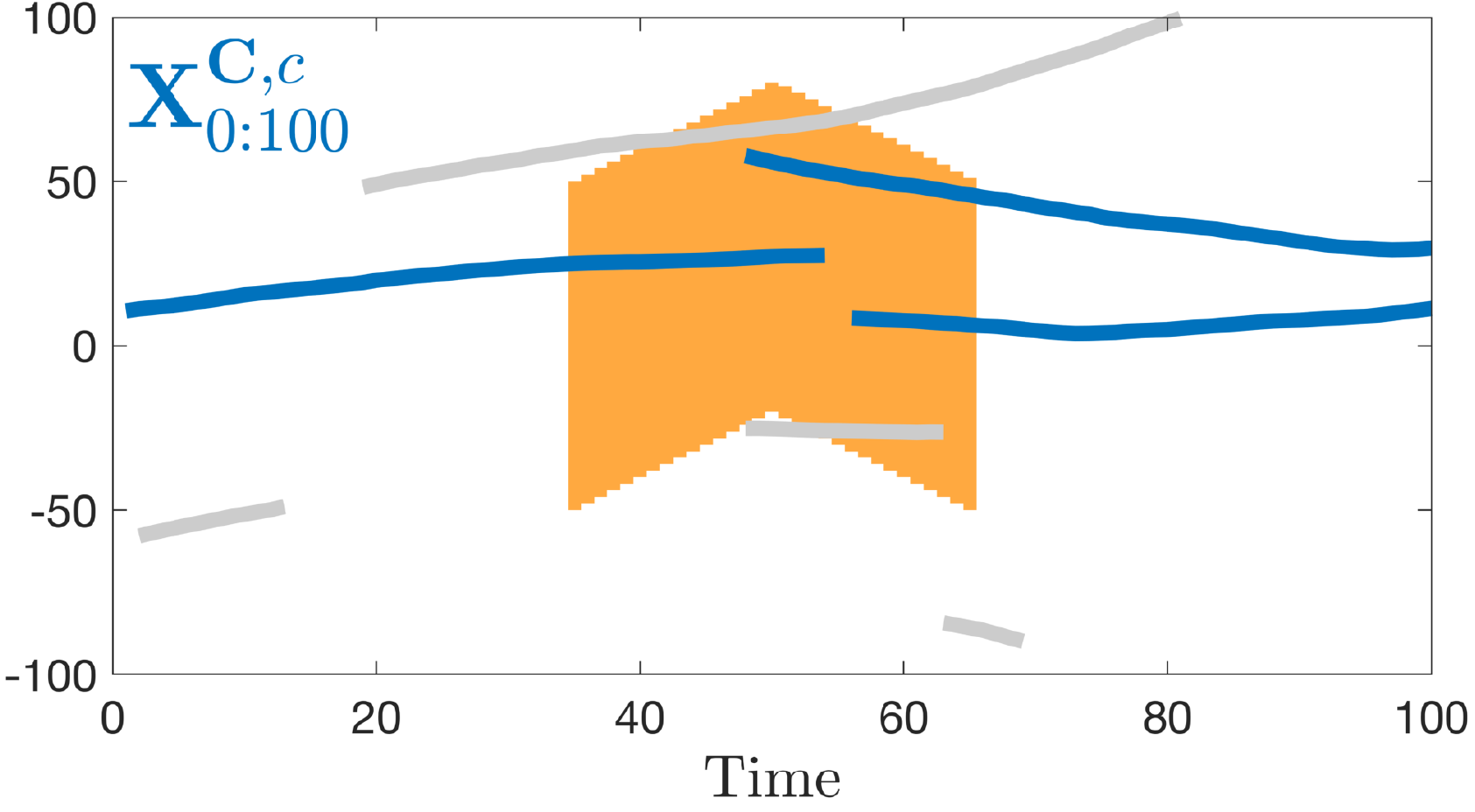}
	\includegraphics[width=0.66\columnwidth]{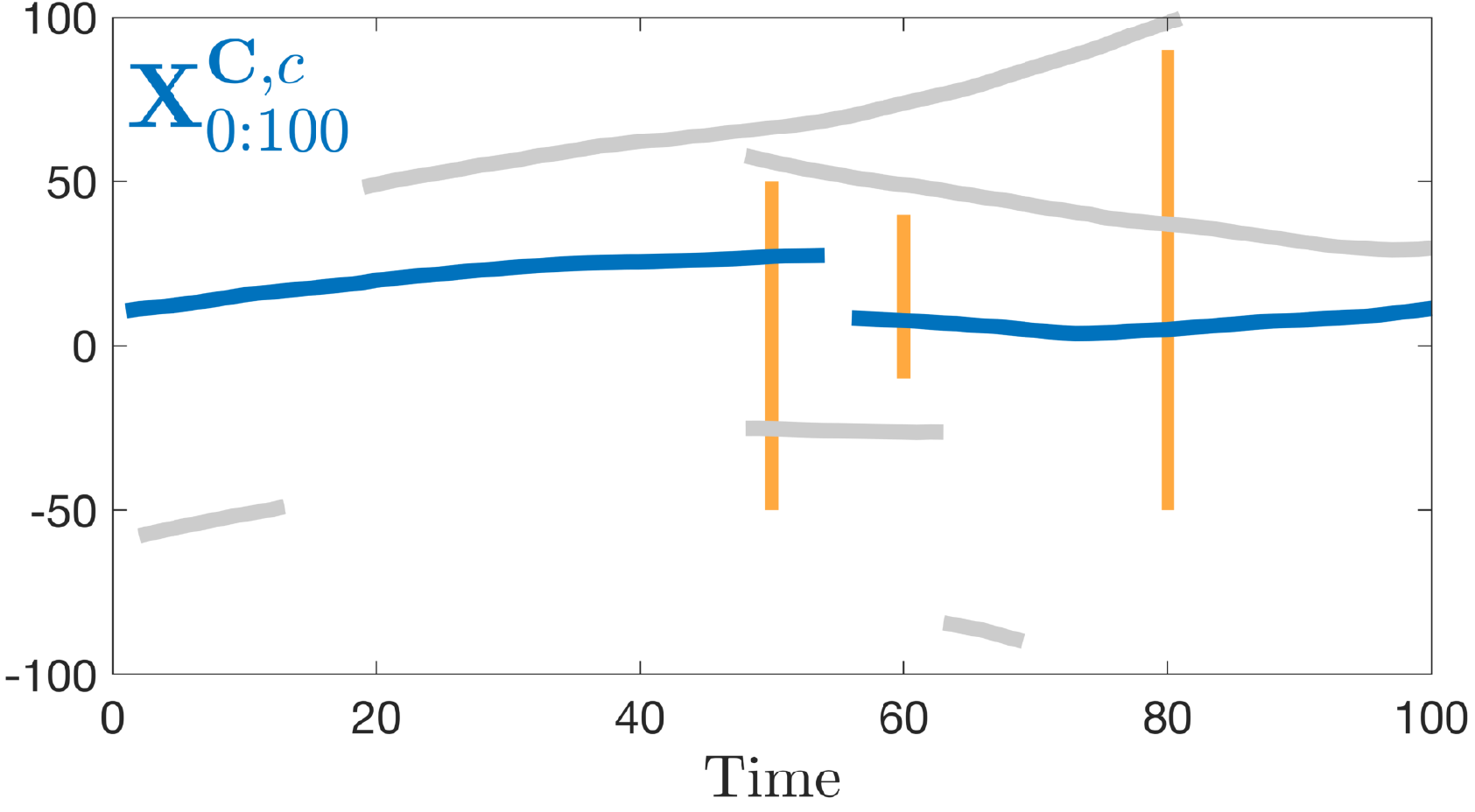}
	
	\caption{Three examples of a set of conjunct trajectory constraints applied to the set of trajectories from Figure~\ref{fig:all_trajectories}. Set of constraints illustrated in orange, trajectories that meet the constraint illustrated in blue, trajectories that do not meet the constraint illustrated in grey. Note that the constraints do not need to have the same spatial regions, nor do they need to be consecutive.}
	\label{fig:conjunct_constraints}
\end{figure*}

\begin{figure*}
	\centering

	\includegraphics[width=0.66\columnwidth]{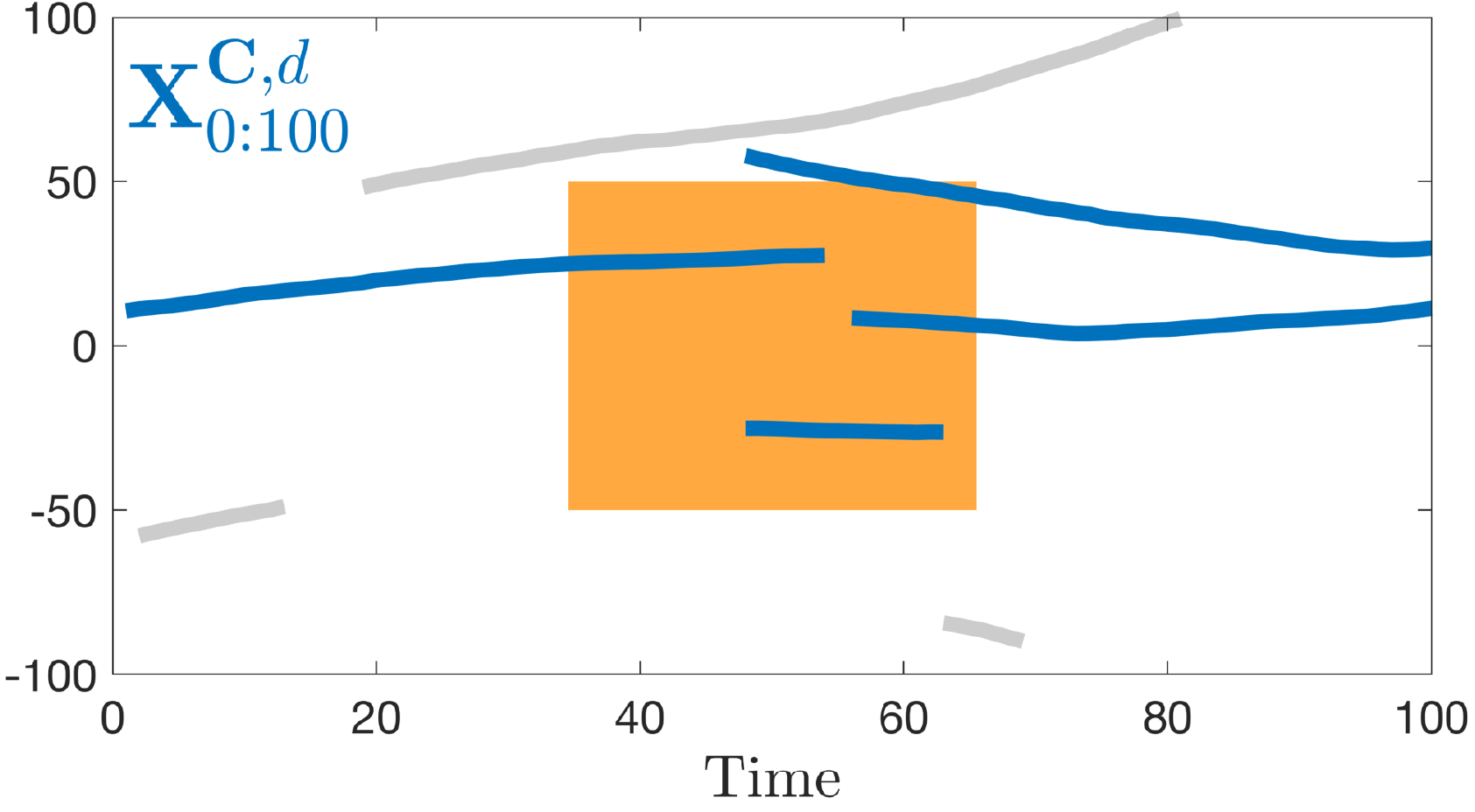}
	\includegraphics[width=0.66\columnwidth]{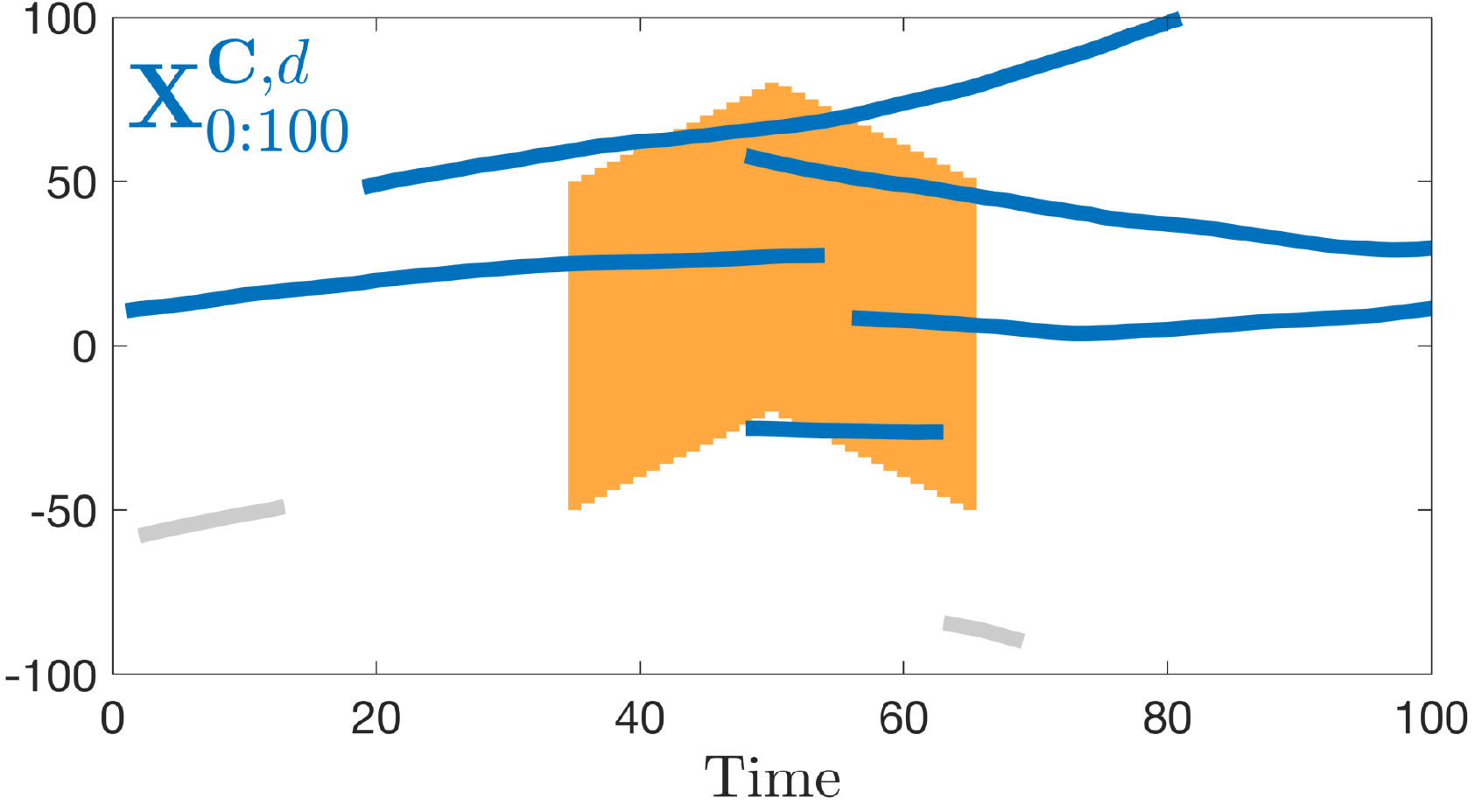}
	\includegraphics[width=0.66\columnwidth]{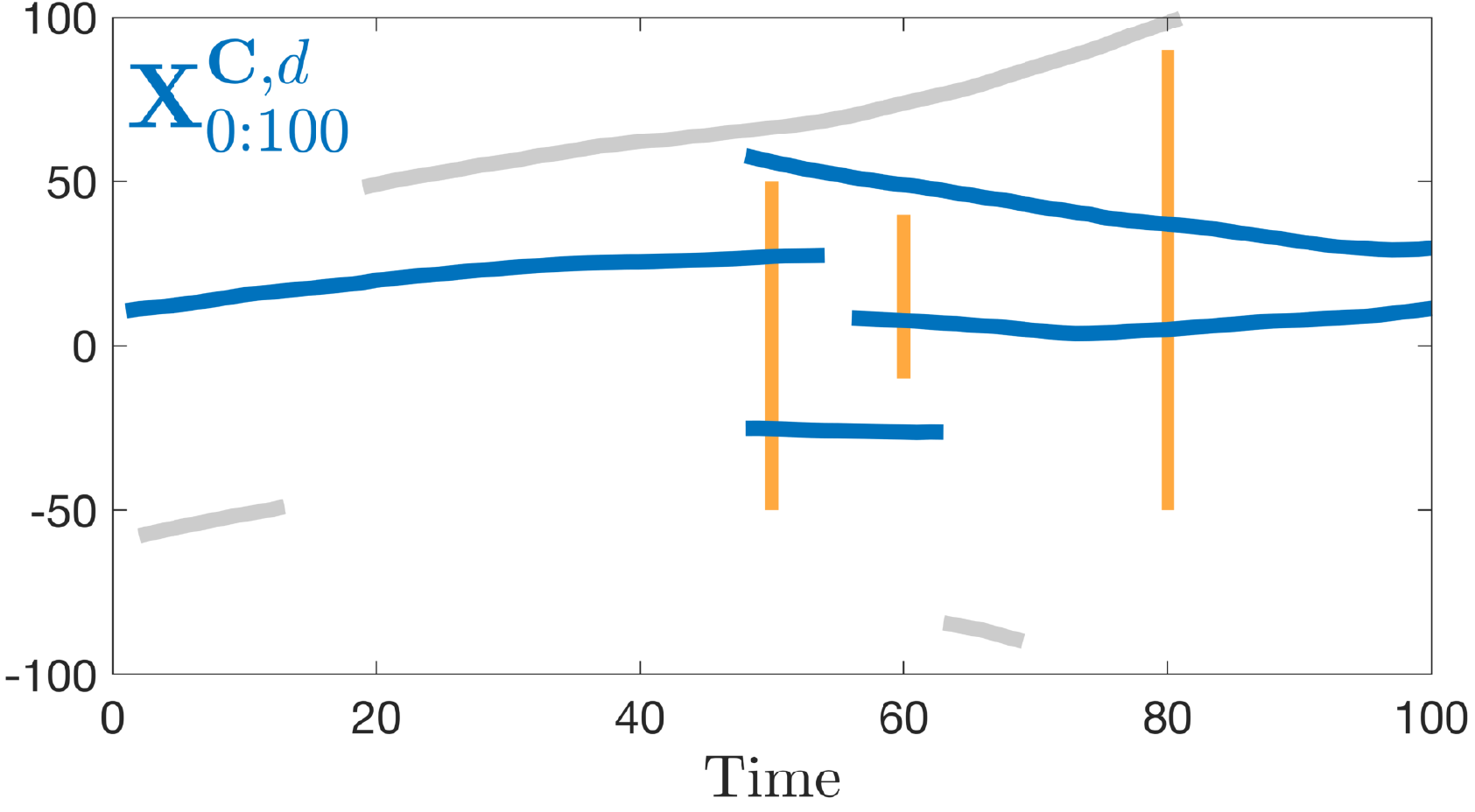}
	
	\caption{Three examples of a set of disjunct trajectory constraints applied to the set of trajectories from Figure~\ref{fig:all_trajectories}. Set of constraints illustrated in orange, trajectories that meet the constraint illustrated in blue, trajectories that do not meet the constraint illustrated in grey. Note that the constraints do not need to have the same spatial regions, nor do they need to be consecutive.}
	\label{fig:disjunct_constraints}
\end{figure*}

%
%
Figure~\ref{fig:conjunct_constraints} and Figure~\ref{fig:disjunct_constraints} show examples of sets of conjunct constraints, and sets of disjunct constraints, applied to the sets of trajectories in Figure~\ref{fig:all_trajectories}.

For any set of constraints $\setconstraint$ we define the set of constraints function for a single trajectory as
\begin{align}
	\tau_{\timeseq{\alpha}{\gamma}}^{\setconstraint} (\traj) = \begin{cases} \left\{ \traj\right\} & \text{if } \traj\in\settraj_{\timeseq{\alpha}{\gamma}}^{\setconstraint}, \\ \emptyset & \text{otherwise.} \end{cases}
\end{align}
For set inputs we have
\begin{align}
	\tau_{\timeseq{\alpha}{\gamma}}^{\setconstraint}(\settraj) & = \begin{cases} \bigcup_{\traj\in\settraj} \tau_{\timeseq{\alpha}{\gamma}}^{\setconstraint}(\traj) & \settraj \neq \emptyset , \\
	\emptyset & \settraj = \emptyset .
	\end{cases}
\end{align}%


\subsection{Relation to previous constraints with trajectories alive in time interval}

The time window constraints presented in \cite{GranstromSXGFW:PMBMtrackersARXIV} mean that the trajectory must be alive at some point in the interval $\timeseq{\eta}{\zeta}$, which corresponds to the following set of trajectory constraints,
\begin{align}
	\left\{ (\eta,\targetStateSpace), \ (\eta+1,\targetStateSpace), \ldots, (\zeta-1,\targetStateSpace),\ (\zeta,\targetStateSpace)\right\} .
\end{align}
There is no spatial constraint, i.e., $\statespaceconstraint=\targetStateSpace$, the temporal constraints are consecutive, and the restriction that the trajectory must be alive in the time interval $\timeseq{\eta}{\zeta}$ means that we have a disjunct set of constraints. A set of spatiotemporal constraints, defined as in Section~\ref{sec:SetOfTrajectoryConstraints}, generalises this in two ways:
\begin{enumerate}
	\item There are spatial constraints, i.e., restrictions on the state of the trajectory at the time step of interest. In other words, it does not only have to be alive, but has to also be located in some part of the state space. Compare Figure~\ref{fig:consecutive_time_constraints}, to Figures~\ref{fig:conjunct_constraints} and \ref{fig:disjunct_constraints}, where spatial constraints are also applied.
	
	\item The trajectory is not restricted to be alive in a time interval, but is restricted to be alive at some point in a set of time steps that is possibly non-consecutive. 
	
\end{enumerate}

%
%


\section{Spatiotemporally constrained PMBM densities}
\label{sec:spatiotemporally_constrained_PMBM_density}

In \cite[Thm. 4]{GranstromSXGFW:PMBMtrackersARXIV} it is shown that, for the standard point target model, given measurement sets in the time interval $\timeseq{\xi}{\chi}$, the density for trajectories in the time interval $\timeseq{\alpha}{\gamma}$ that are alive at any point in the time interval $\timeseq{\eta}{\zeta}$, $\timeset{\eta}{\zeta}\subseteq\timeset{\alpha}{\gamma}$, 
	\begin{align}
		f_{\timeseq{\alpha}{\gamma} | \timeseq{\xi}{\chi} }^{\timeseq{\eta}{\zeta}}\left(\settraj_{\timeseq{\alpha}{\gamma}}^{\timeseq{\eta}{\zeta}}\right), \label{eq:General_PMBM_density}
\end{align}
is exactly \pmbm for any time intervals $\timeseq{\xi}{\chi}$ and $\timeseq{\alpha}{\gamma}$, and any interval $\timeseq{\eta}{\zeta}$ such that $\timeset{\eta}{\zeta}\subseteq\timeset{\alpha}{\gamma}$. In this section we generalise this property in the following theorem.
\begin{theorem}
	For the standard point target model, given measurement sets in the time interval $\timeseq{\xi}{\chi}$, the density for trajectories in the time interval $\timeseq{\alpha}{\gamma}$ that satisfy a set of constraints $\setconstraint$, 
	\begin{align}
		f_{\timeseq{\alpha}{\gamma} | \timeseq{\xi}{\chi} }^{\setconstraint}\left(\settraj_{\timeseq{\alpha}{\gamma}}^{\setconstraint}\right), \label{eq:General_PMBM_density}
	\end{align}
	is exactly \pmbm for any time intervals $\timeseq{\xi}{\chi}$ and $\timeseq{\alpha}{\gamma}$, and any set of constraints $\setconstraint$ (conjunct or disjunct). 
\end{theorem}

In the following subsections we present Lemmas that give the Bernoulli and \ppp parameters that correspond to the Theorem. We consider three cases: to keep things easy initially, we begin with the special case where we have a single spatiotemporal constraint, and then we consider conjunct constraints and disjunct constraints, respectively. The full proofs are too lengthy to be included here; some details are provided in the Appendix.


\subsection{Single spatiotemporal constraint}

\begin{lemma}\label{lem:SingleConstraintTrajectoryDensity}
Let $p(\traj) = p(\stseq_{\timeseq{\tb}{\td}}|\tb,\td)P(\tb,\td)$ be a trajectory density, and let $\trajconstraint = (\timeconstraint, \ \statespaceconstraint)$ be a single spatiotemporal constraint. The constrained trajectory density is then
\begin{subequations}
\begin{align}
	p^{\trajconstraint}(\traj) =& p^{\trajconstraint}(\stseq_{\timeseq{\tb}{\td}}|\tb,\td)P^{\trajconstraint}(\tb,\td) \\
	p^{\trajconstraint}(\stseq_{\timeseq{\tb}{\td}}|\tb,\td) =& \begin{cases} \frac{p(\stseq_{\timeseq{\tb}{\td}}|\tb,\td)}{\Pr(x_{\timeconstraint}\in\statespaceconstraint | \timeconstraint\in\timeset{\tb}{\td})}  & \text{if } x_{\timeconstraint}\in\statespaceconstraint \\ 0 & \text{otherwise}\end{cases} \\
	P^{\trajconstraint}(\tb,\td) = & \begin{cases} \frac{P(\tb,\td)}{\Pr(\timeconstraint\in\timeset{\tb}{\td})} & \text{if }k\in\timeset{\tb}{\td} \\ 0 & \text{otherwise} \end{cases}
\end{align}%
\label{eq:SingleConstrainedTrajectoryDensity}%
\end{subequations}
where the probability that the trajectory is alive at the time of the constraint is
\begin{align}
	\Pr(\timeconstraint\in\timeset{\tb}{\td}) = \sum_{\substack{\tb,\td: \\ \timeconstraint\in\timeset{\tb}{\td}}} P(\tb,\td)
\end{align}
and the probability that the state is located in the constraint region is
\begin{align}
	\Pr\left(x_{\timeconstraint}\in\statespaceconstraint \left| \timeconstraint\in\timeset{\tb}{\td}\right.\right) 
	& = \int_{\statespaceconstraint} p(\stseq_{\timeconstraint} | \tb,\td) \diff x_{\timeconstraint}
\end{align}
Note that $\int p^{\trajconstraint}(\traj) \diff \traj = 1$.
\end{lemma}

\begin{lemma}
Let $f(\settraj)$ be a trajectory Bernoulli density with parameters $r$ and $p(\traj)$, and let $\trajconstraint = (\timeconstraint, \ \statespaceconstraint)$ be a single spatiotemporal constraint. The constrained set of trajectories density is a trajectory Bernoulli density with probability of existence
\begin{align}
	r^{\trajconstraint} 
	= & r \Pr(x_{\timeconstraint}\in\statespaceconstraint | \timeconstraint\in\timeset{\tb}{\td}) \Pr(\timeconstraint\in\timeset{\tb}{\td})
\end{align}
and constrained trajectory density $p^{\trajconstraint}(\cdot)$ given by Lemma~\ref{lem:SingleConstraintTrajectoryDensity}; $\Pr(x_{\timeconstraint}\in\statespaceconstraint | \timeconstraint\in\timeset{\tb}{\td})$, $\Pr(\timeconstraint\in\timeset{\tb}{\td})$ are given by Lemma~\ref{lem:SingleConstraintTrajectoryDensity}. 
\end{lemma}

\begin{lemma}
Let $f(\settraj)$ be a Poisson Point Process (\ppp) with intensity $\lambda(\traj) = \mu p(\traj)$, where $\mu>0$ and $p(\traj)$ is a single trajectory density, and let $\trajconstraint = (\timeconstraint, \ \statespaceconstraint)$ be a single spatiotemporal constraint. 
The constrained set of trajectories density is a \ppp density with intensity
\begin{align}
	\lambda^{\trajconstraint}(\traj) = \mu^{\trajconstraint} p^{\trajconstraint}(\traj)
\end{align}
where
\begin{align}
	\mu^{\trajconstraint} = \mu \Pr(x_{\timeconstraint}\in\statespaceconstraint | \timeconstraint\in\timeset{\tb}{\td}) \Pr(\timeconstraint\in\timeset{\tb}{\td}),
\end{align}
and $p^{\trajconstraint}(\traj)$, $\Pr(x_{\timeconstraint}\in\statespaceconstraint | \timeconstraint\in\timeset{\tb}{\td})$, and $\Pr(\timeconstraint\in\timeset{\tb}{\td})$ are given by Lemma~\ref{lem:SingleConstraintTrajectoryDensity}.
\end{lemma}


\subsection{Set of conjunct constraints}
\begin{lemma}\label{lem:ConjunctConstraintsTrajectoryDensity}
Let $p(\traj) = p(\stseq_{\timeseq{\tb}{\td}}|\tb,\td)P(\tb,\td)$ be a trajectory density, and let $\setconstraint$ be a set of conjunct constraints. The constrained set of trajectories density is then
\begin{subequations}
\begin{align}
	& p^{\setconstraint}(\traj) = p^{\setconstraint}(\stseq_{\timeseq{\tb}{\td}}|\tb,\td)P^{\setconstraint}(\tb,\td) \\
	& p^{\setconstraint}(\stseq_{\timeseq{\tb}{\td}}|\tb,\td) \nonumber \\
	& \quad = \begin{cases} \frac{p(\stseq_{\timeseq{\tb}{\td}}|\tb,\td)}{\Pr(x_{\timeconstraint_{i}} \in \statespaceconstraint_{i}, \  \forall i \in \mathbb{I}_{\tb}^{\td} | \settimeconstraint_{\tb}^{\td} \neq \emptyset )}  & \text{if } x_{\timeconstraint_{i}} \in \statespaceconstraint_{i}, \  \forall i \in \mathbb{I}_{\tb}^{\td} \\ 0 & \text{otherwise}\end{cases} \\
	& P^{\setconstraint}(\tb,\td) =  \begin{cases} \frac{P(\tb,\td)}{\Pr(\settimeconstraint_{\tb}^{\td} \neq \emptyset)} & \text{if } \settimeconstraint_{\tb}^{\td} \neq \emptyset \\ 0 & \text{otherwise} \end{cases}
\end{align}%
\label{eq:ConjunctConstraintsTrajectoryDensity}%
\end{subequations}
where the probability that at least one of the constraints occur in the time interval $\timeseq{\tb}{\td}$ is
\begin{align}
	\Pr( \settimeconstraint_{\tb}^{\td} \neq \emptyset ) = \sum_{\substack{\tb,\td: \\ \settimeconstraint_{\tb}^{\td} \neq \emptyset }} P(\tb,\td)
\end{align}
and the probability that the state is inside the constraint regions is
\begin{align}
	& \Pr(x_{\timeconstraint_{i}} \in \statespaceconstraint_{i}, \  \forall i \in \mathbb{I}_{\tb}^{\td} | \settimeconstraint_{\tb}^{\td} \neq \emptyset ) = \int_{\times_{i\in\mathbb{I}_{\tb}^{\td}}\statespaceconstraint_{i}} p(\stseq_{\settimeconstraint_{\tb}^{\td}} |\tb,\td) \diff \stseq_{\settimeconstraint_{\tb}^{\td}} 
	&
\end{align}
Note that $\int p^{\setconstraint}(\traj) \diff \traj = 1$.
\end{lemma}

\begin{lemma}
Let $f(\settraj)$ be a trajectory Bernoulli density with parameters $r$ and $p(\traj) = p(\stseq_{\timeseq{\tb}{\td}}|\tb,\td)P(\tb,\td)$, and let $\setconstraint$ be a set of conjunct constraints. The constrained set of trajectories density is a Bernoulli density with probability of existence
\begin{align}
	r^{\setconstraint} 
	= & r \Pr(x_{\timeconstraint_{i}} \in \statespaceconstraint_{i}, \  \forall i \in \mathbb{I}_{\tb}^{\td} | \settimeconstraint_{\tb}^{\td} \neq \emptyset ) \Pr( \settimeconstraint_{\tb}^{\td} \neq \emptyset )
\end{align}
and trajectory density $p^{\setconstraint}(\cdot)$ given by Lemma~\ref{lem:ConjunctConstraintsTrajectoryDensity}; $ \Pr(x_{\timeconstraint_{i}} \in \statespaceconstraint_{i}, \  \forall i \in \mathbb{I}_{\tb}^{\td} | \settimeconstraint_{\tb}^{\td} \neq \emptyset )$ and $\Pr( \settimeconstraint_{\tb}^{\td} \neq \emptyset )$ are given by Lemma~\ref{lem:ConjunctConstraintsTrajectoryDensity}.
\end{lemma}

\begin{lemma}
Let $f(\settraj)$ be a Poisson Point Process (\ppp) with intensity $\lambda(\traj) = \mu p(\traj)$, and let $\setconstraint$ be a set of conjunct constraints. The constrained set of trajectories density is a \ppp density with intensity
\begin{align}
	\lambda^{\setconstraint}(\traj) = \mu^{\setconstraint} p^{\setconstraint}(\traj)
\end{align}
where
\begin{align}
	\mu^{\setconstraint} = \mu \Pr(x_{\timeconstraint_{i}} \in \statespaceconstraint_{i}, \  \forall i \in \mathbb{I}_{\tb}^{\td} | \settimeconstraint_{\tb}^{\td} \neq \emptyset ) \Pr( \settimeconstraint_{\tb}^{\td} \neq \emptyset ),
\end{align}
and $p^{\setconstraint}(\cdot)$, $\Pr(x_{\timeconstraint_{i}} \in \statespaceconstraint_{i}, \  \forall i \in \mathbb{I}_{\tb}^{\td} | \settimeconstraint_{\tb}^{\td} \neq \emptyset )$, and $\Pr( \settimeconstraint_{\tb}^{\td} \neq \emptyset )$ are given by Lemma~\ref{lem:ConjunctConstraintsTrajectoryDensity}.
\end{lemma}


\subsection{Set of disjunct constraints}

\begin{lemma}\label{lem:DisjunctConstraintsTrajectoryDensity}
Let $p(\traj) = p(\stseq_{\timeseq{\tb}{\td}}|\tb,\td)P(\tb,\td)$ be a trajectory density, and let $\setconstraint$ be a set of disjunct constraints. The constrained trajectory density is then
\begin{subequations}
\begin{align}
	& p^{\setconstraint}(\traj) = p^{\setconstraint}(\stseq_{\timeseq{\tb}{\td}}|\tb,\td)P^{\setconstraint}(\tb,\td) \\
	& P^{\setconstraint}(\tb,\td) =  \begin{cases} \frac{P(\tb,\td)}{\Pr(\settimeconstraint_{\tb}^{\td} \neq \emptyset)} & \text{if } \settimeconstraint_{\tb}^{\td} \neq \emptyset \\ 0 & \text{otherwise} \end{cases} \\
	& p^{\setconstraint}(\stseq_{\timeseq{\tb}{\td}}|\tb,\td) \\
	& \quad = \begin{cases} 0 & x_{\timeconstraint_{i}} \in \statespaceconstraint_{i}^{\complement}, \ \forall i \in \mathbb{I}_{\tb}^{\td}  \\ \sum_{\substack{\mathbb{I}^{i} \uplus \mathbb{I}^{o} = \mathbb{I}_{\tb}^{\td}: \\ \mathbb{I}^{i} \neq \emptyset}} w^{\mathbb{I}^{i} , \mathbb{I}^{o}} p^{\mathbb{I}^{i},\mathbb{I}^{o}}(\stseq_{\timeseq{\tb}{\td}})  & \text{otherwise } \end{cases} \nonumber \\
	& p^{\mathbb{I}^{i},\mathbb{I}^{o}}(\stseq_{\timeseq{\tb}{\td}}) \\
	& \quad = \begin{cases} \frac{p(\stseq_{\timeseq{\tb}{\td}} | \tb , \td) }{ \tilde{w}^{\mathbb{I}^{i},\mathbb{I}^{o}}} & x_{\timeconstraint_{i}}\in\statespaceconstraint_{i}, \forall i\in\mathbb{I}^{i} \text{ and } x_{\timeconstraint_{j}}\in\statespaceconstraint_{j}^{\complement}, \forall j\in\mathbb{I}^{o} \\ 0 & \text{otherwise} \end{cases} \nonumber \\
	& w^{\mathbb{I}^{i},\mathbb{I}^{o}} = \frac{ \tilde{w}^{\mathbb{I}^{i},\mathbb{I}^{o}} }{ \sum_{\substack{\mathbb{I}^{i} \uplus \mathbb{I}^{o} = \mathbb{I}_{\tb}^{\td}: \\ \mathbb{I}^{i} \neq \emptyset}} \tilde{w}^{\mathbb{I}^{i},\mathbb{I}^{o}} } \\
	& \tilde{w}^{\mathbb{I}^{i},\mathbb{I}^{o}} = \int_{\times_{i\in\mathbb{I}^{i}} \statespaceconstraint_{i}} \int_{\times_{j\in\mathbb{I}^{o}} \statespaceconstraint_{j}^{\complement}} p(\stseq_{\settimeconstraint^{i} \cup \settimeconstraint^{o}} | \tb , \td)  \diff \stseq_{\settimeconstraint^{i}} \diff \stseq_{\settimeconstraint^{o}}
\end{align}%
\label{eq:DisjunctConstraintsTrajectoryDensity}%
\end{subequations}
where $\mathbb{I}^{i}$ and $\mathbb{I}^{o}$ are index sets corresponding to the spatial constraints that are satisfied and not, respectively, $\settimeconstraint^{i}$ and $\settimeconstraint^{o}$ are the corresponding time steps, the probability that at least one of the constraints occur in the time interval $\timeseq{\tb}{\td}$ is
\begin{align}
	\Pr( \settimeconstraint_{\tb}^{\td} \neq \emptyset ) = \sum_{\substack{\tb,\td: \\ \settimeconstraint_{\tb}^{\td} \neq \emptyset }} P(\tb,\td)
\end{align}
and the probability that the state is inside at least one of the constraint regions is
equivalent to one minus the probability that the state is inside non of the constraint regions, 
\begin{align}
	& \Pr( \exists i \in \mathbb{I}_{\tb}^{\td} : x_{\timeconstraint_{i}} \in \statespaceconstraint_{i} | \settimeconstraint_{\tb}^{\td} \neq \emptyset ) \nonumber \\
	& = 1 - \Pr( x_{\timeconstraint_{i}} \in \statespaceconstraint_{i}^{\complement}, \ \forall i \in \mathbb{I}_{\tb}^{\td} | \settimeconstraint_{\tb}^{\td} \neq \emptyset )
\end{align}
where
\begin{align}
	& \Pr(x_{\timeconstraint_{i}} \in \statespaceconstraint_{i}^{\complement}, \  \forall i \in \mathbb{I}_{\tb}^{\td} | \settimeconstraint_{\tb}^{\td} \neq \emptyset ) \nonumber \\
	& \quad = \int_{\times_{i\in\mathbb{I}_{\tb}^{\td}}\statespaceconstraint_{i}^{\complement}} p(\stseq_{\settimeconstraint_{\tb}^{\td}} | \tb,\td) \diff \stseq_{\settimeconstraint_{\tb}^{\td}} 
\end{align}
Note that $\int p^{\setconstraint}(\traj) \diff \traj = 1$.
\end{lemma}

\begin{lemma}
Let $f(\settraj)$ be a trajectory Bernoulli density with parameters $r$ and $p(\traj) = p(\stseq_{\timeseq{\tb}{\td}}|\tb,\td)P(\tb,\td)$, and let $\setconstraint$ be a set of disjunct constraints. The constrained set of trajectories density is a Bernoulli density with probability of existence
\begin{align}
	r^{\setconstraint} 
	= & r \left( 1 - \Pr( x_{\timeconstraint_{i}} \in \statespaceconstraint_{i}^{\complement}, \ \forall i \in \mathbb{I}_{\tb}^{\td} | \settimeconstraint_{\tb}^{\td} \neq \emptyset ) \right) \Pr( \settimeconstraint_{\tb}^{\td} \neq \emptyset )
\end{align}
and trajectory density $p^{\setconstraint}(\cdot)$ given by Lemma~\ref{lem:DisjunctConstraintsTrajectoryDensity}; $\Pr( x_{\timeconstraint_{i}} \in \statespaceconstraint_{i}^{\complement}, \ \forall i \in \mathbb{I}_{\tb}^{\td} | \settimeconstraint_{\tb}^{\td} \neq \emptyset )$ and $\Pr( \settimeconstraint_{\tb}^{\td} \neq \emptyset )$ are given by  Lemma~\ref{lem:DisjunctConstraintsTrajectoryDensity}.
\end{lemma}

\begin{lemma}
Let $f(\settraj)$ be a Poisson Point Process (\ppp) with intensity $\lambda(\traj) = \mu p(\stseq_{\timeseq{\tb}{\td}}|\tb,\td)P(\tb,\td)$, and let $\setconstraint$ be a set of disjunct constraints. The constrained set of trajectories density is a \ppp with intensity
\begin{align}
	\lambda^{\setconstraint}(\traj) = \mu^{\setconstraint} p^{\setconstraint}(\traj)
\end{align}
where
\begin{align}
	\mu^{\setconstraint} = \mu\left( 1 - \Pr( x_{\timeconstraint_{i}} \in \statespaceconstraint_{i}^{\complement}, \ \forall i \in \mathbb{I}_{\tb}^{\td} | \settimeconstraint_{\tb}^{\td} \neq \emptyset ) \right) \Pr( \settimeconstraint_{\tb}^{\td} \neq \emptyset ),
\end{align}
and $p^{\setconstraint}(\cdot)$, $\Pr( x_{\timeconstraint_{i}} \in \statespaceconstraint_{i}^{\complement}, \ \forall i \in \mathbb{I}_{\tb}^{\td} | \settimeconstraint_{\tb}^{\td} \neq \emptyset )$ and $\Pr( \settimeconstraint_{\tb}^{\td} \neq \emptyset )$ are given by Lemma~\ref{lem:DisjunctConstraintsTrajectoryDensity}.
\end{lemma}


\section{Examples}
\label{sec:examples}

\begin{figure*}
	\includegraphics[width=0.5\textwidth]{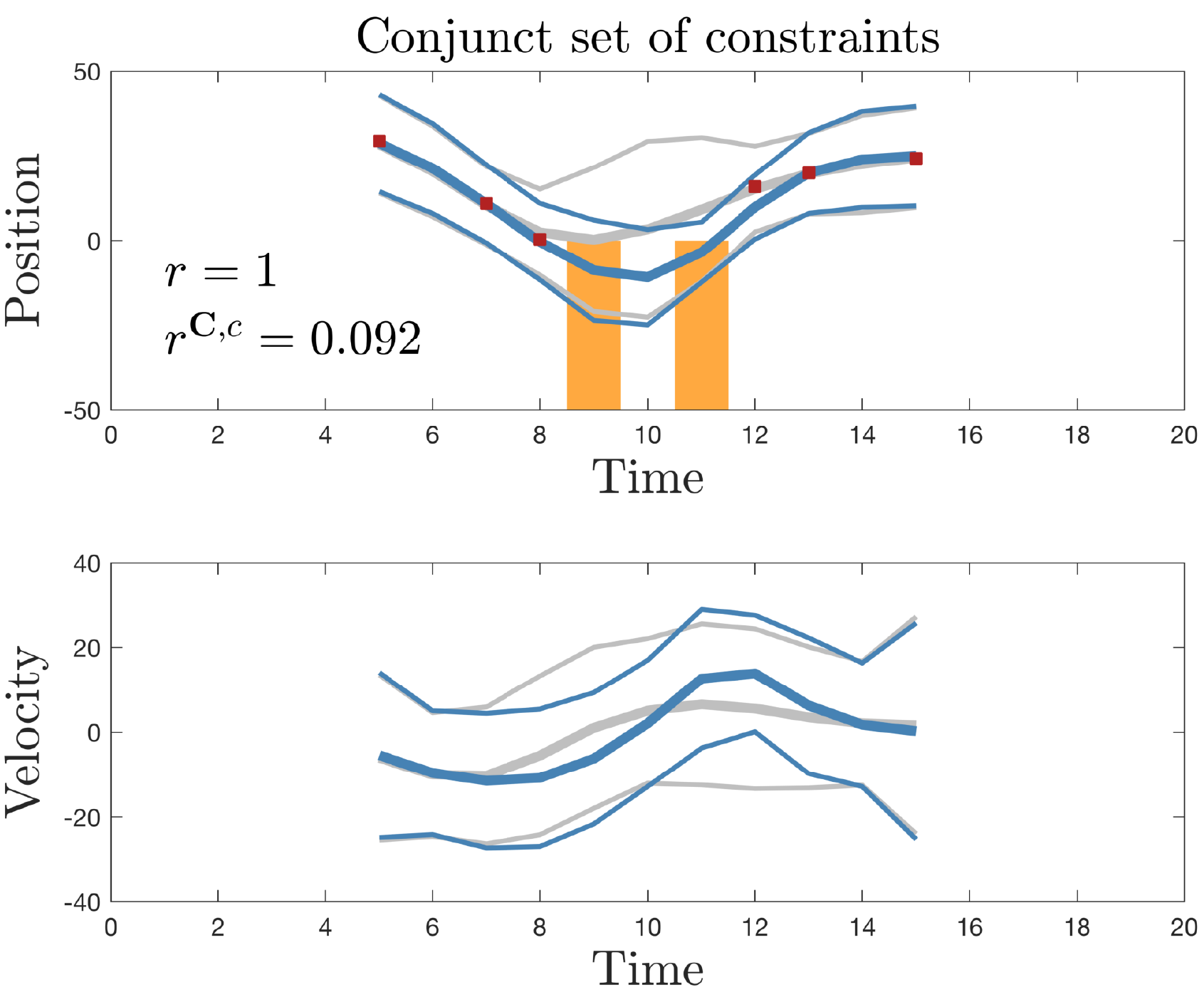}
	\includegraphics[width=0.5\textwidth]{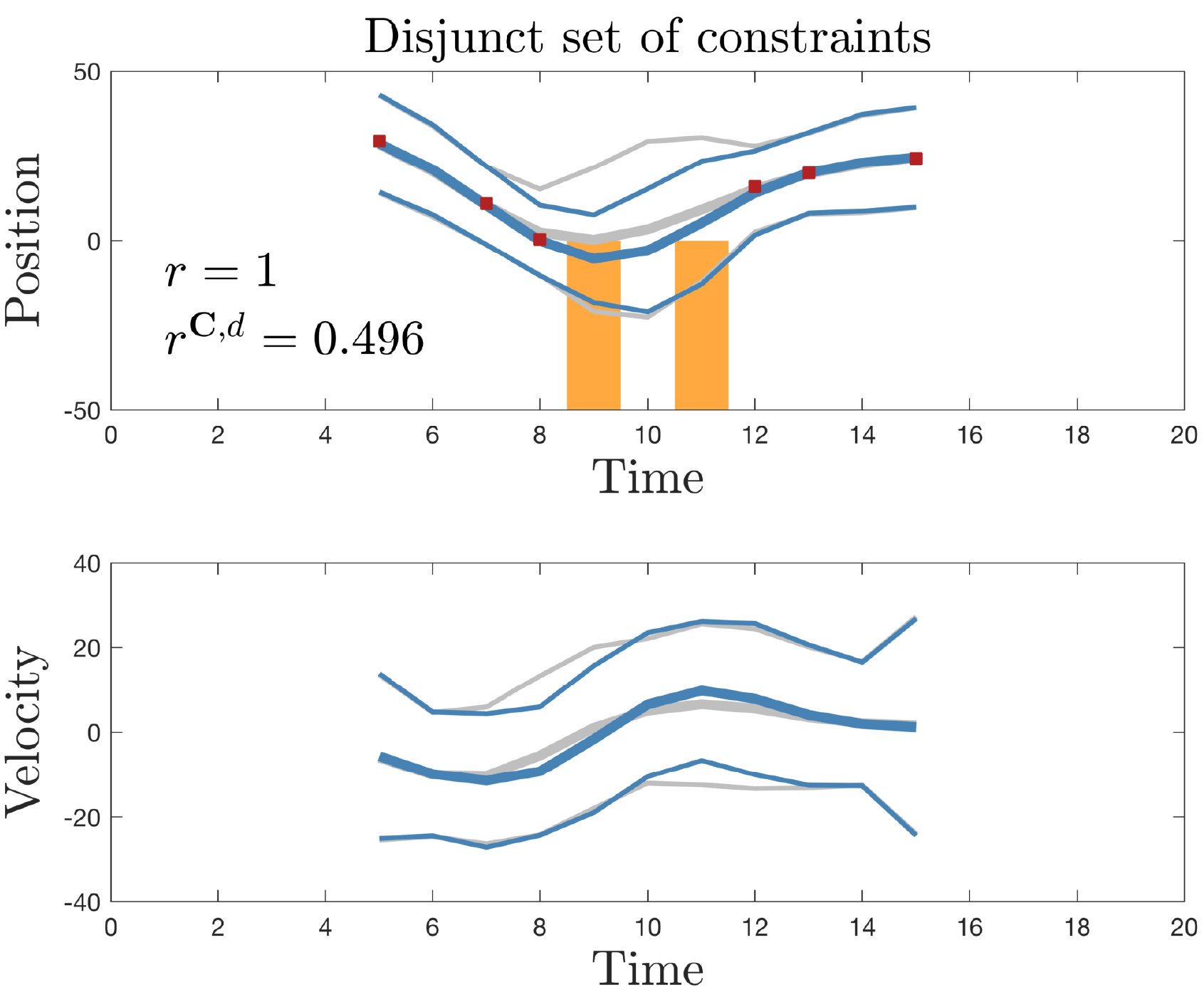}
	\caption{Example of constrained Bernoulli density. Associated measurements (red squares); spatiotemporal constraints (orange); unconstrained Gaussian state sequence density (gray) showing mean and mean $\pm$ three standard deviations; Gaussian approximation of constrained state sequence density (blue) showing mean and mean $\pm$ three standard deviations.}
	\label{fig:ConstraintsExample_1D}
\end{figure*}

A scenario where the target state consists of 1D position and velocity was simulated. In Figure~\ref{fig:ConstraintsExample_1D} we show example results where a set of constraints was applied to a Bernoulli density, both conjunct constrains and disjunct constraints. The constrained state sequence densities are different depending on if the set of constraints is conjunct or disjunct, and the difference in constrained probability of existence $r^{\mathbf{C}}$ is quite large. The constrained state densities were computed using Monte Carlo approximation of the integrals. 

\section{Concluding remarks}
\label{sec:conclusion}

In this paper we introduced spatiotemporal constraints for sets of trajectories, and applied them to Poisson Multi-Bernoulli Mixture (\pmbm) densities. We showed that if we have a \pmbm set of trajectories density, then the density for trajectories that satisfy a set of constraints is also \pmbm. From this it follows that a constrained Bernoulli is also a Bernoulli, and a constrained \ppp is also a \ppp. An important topic for future work is developing computationally efficient methods to compute the constrained trajectory densities.





\appendix

In what follows, for the sake of brevity, we drop the sub-indexing w.r.t. the time intervals $\timeseq{\xi}{\chi}$ and $\timeseq{\alpha}{\gamma}$. From \cite[Thm. 11]{GarciaFernandezSM:2019} we know that we can formulate a transition density from the set $\settraj$ to the set $\settraj^{\setconstraint}=\tau^{\setconstraint}(\settraj)$ as
\begin{align}
	\delta_{\tau^{\setconstraint}(\settraj)}\left(\settraj^{\setconstraint}\right).
\end{align}
If $f(\settraj)$ is a \pmbm density, then, if $\setY=\tau^{\setconstraint}(\settraj)$, the density for $\setY$ is given by \cite[Lem. 1]{GranstromSXGFW:PMBMtrackersARXIV}
\begin{subequations}
	\begin{align}
		g(\setY) & = \int \delta_{ \tau^{\setconstraint}(\settraj)}(\setY) f(\settraj) \delta \settraj  \\
		&= \sum_{\substack{\left(\uplus_{i\in\trackTable} \setY^{i} \right) \uplus \setY^{\rm u} = \setY}} \int \delta_{ \tau^{\setconstraint}\left( \settraj^{\rm u}\right)}(\setY^{\rm u})  f^{\rm u}(\settraj^{\rm u}) \delta \settraj^{\rm u} \nonumber \\
		& \quad \times \sum_{\assoc\in\assocspace} w_{\assoc} \prod_{i\in\mathbb{T}} \int \delta_{ \tau^{\setconstraint}\left( \settraj^i\right)}(\setY^i)  f^{i,a^i}(\settraj^i) \delta \settraj^i 
	\end{align}
\end{subequations}
To show that $g(\setY)$ is \pmbm, we need to show that $\int \delta_{ \tau^{\setconstraint}\left( \settraj^{\rm u}\right)}(\setY^{\rm u})  f^{\rm u}(\settraj^{\rm u}) \delta \settraj^{\rm u}$ is a \ppp, and that $\int \delta_{ \tau^{\setconstraint}\left( \settraj^i\right)}(\setY^i)  f^{i,a^i}(\settraj^i) \delta \settraj^i$ is a Bernoulli density. 

For a Bernoulli density, we have (indexing omitted for the sake of brevity)
\begin{align}
	g(\setY) = & \int \delta_{\tau^{\setconstraint}(\settraj)}(\setY) f(\settraj)\delta\settraj \\
	= & (1-r) \delta_{\emptyset}(\setY) + r \int p(\traj)  \delta_{\tau^{\setconstraint}(\{\traj\})}(\setY) \diff \traj 
\end{align}
For a \ppp density with intensity $\lambda(\cdot)$, from \cite[p. 99, Rmk. 12]{Mahler:2014} it follows that the \ppp can be divided into two disjoint and independent \ppp subsets, with intensity $\lambda^{\setconstraint}(\cdot)$ for trajectories that meet the constraints (conjunct or disjunct) and $\lambda^{\not\setconstraint}(\cdot)$ for trajectories that do not meet the constraints. The constrained intensity is given by
\begin{align}
	\lambda^{\setconstraint}( Y ) = \int \delta_{\tau^{\setconstraint}(\{\traj\})}(\{Y\}) \lambda(\traj)\diff\traj
\end{align}
For both the Bernoulli and the \ppp we have integrals of the type
\begin{align}
	& \int p(\traj)  \delta_{\tau^{\setconstraint}(\{\traj\})}(\setY) \diff \traj \\
	= & \sum_{\tb,\td} \int_{\targetStateSpace^{\tlen}} p(\stseq_{\timeseq{\tb}{\td}}|\tb,\td)P(\tb,\td) \delta_{\tau^{\setconstraint}(\{\traj\})}(\setY) \diff \stseq_{\timeseq{\tb}{\td}}\\
	= & \sum_{\tb,\td} P(\tb,\td)\int_{\targetStateSpace^{\tlen}} p(\stseq_{\timeseq{\tb}{\td}}|\tb,\td)\delta_{\tau^{\setconstraint}(\{\traj\})}(\setY) \diff \stseq_{\timeseq{\tb}{\td}}
\end{align}
For a set of constraints, we can express the integral as follows,
\begin{align}
	\int & p(\stseq_{\timeseq{\tb}{\td}}|\tb,\td) \delta_{\tau^{\setconstraint}(\{\traj\})}(\setY)  \diff \stseq_{\timeseq{\tb}{\td}} \nonumber \\
	& = \sum_{\mathbb{I}^{i} \uplus \mathbb{I}^{o} = \mathbb{I}_{\tb}^{\td}} \int_{\times_{i\in\mathbb{I}^{i}} \statespaceconstraint_{i}} \int_{\times_{j\in\mathbb{I}^{o}} \statespaceconstraint_{j}^{\complement}}  \\ 
	& \qquad p(\stseq_{\settimeconstraint^{i} \cup \settimeconstraint^{o}} | \tb , \td) \delta_{\tau^{\setconstraint}(\{\traj\})}(\setY) \diff \stseq_{\settimeconstraint^{i}} \diff \stseq_{\settimeconstraint^{o}} \nonumber 
\end{align}
where $\mathbb{I}^{i}$ and $\mathbb{I}^{o}$ are index sets corresponding to the spatial constraints that are satisfied and not, respectively, $\settimeconstraint^{i}$ and $\settimeconstraint^{o}$ are the corresponding time steps. We have expressed the integral as the sum of each way to combine the different spatial constraints, either inside the constraint region $\statespaceconstraint$, or outside it (i.e., inside its complement $\statespaceconstraint^{\complement}$). Depending on if the set of constraints is conjunct or disjunct, this integral will evaluate differently.

\subsection{Conjunct constraints}

When the constraints are conjunct, the state has to be in every constraint region for the time steps that the trajectory is alive:
\begin{align}
	& \int p(\stseq_{\timeseq{\tb}{\td}}) \delta_{\tau^{\setconstraint}(\{\traj\})}(\setY) \diff \stseq_{\timeseq{\tb}{\td}} \nonumber\\
	& = \sum_{\substack{\mathbb{I}^{i} \uplus \mathbb{I}^{o} = \mathbb{I}_{\tb}^{\td}: \\ \mathbb{I}^{o}\neq\emptyset}}  \int_{\times_{i\in\mathbb{I}^{i}} \statespaceconstraint_{i}} \int_{\times_{j\in\mathbb{I}^{o}} \statespaceconstraint_{j}^{\complement}}  \\ 
	& \qquad p(\stseq_{\settimeconstraint^{i} \cup \settimeconstraint^{o}} | \tb , \td) \delta_{\emptyset}(\setY) \diff \stseq_{\settimeconstraint^{i}} \diff \stseq_{\settimeconstraint^{o}} \nonumber \\
	& \quad + \sum_{\substack{\mathbb{I}^{i} \uplus \mathbb{I}^{o} = \mathbb{I}_{\tb}^{\td}: \\ \mathbb{I}^{o}=\emptyset}} \int_{\times_{i\in\mathbb{I}^{i}} \statespaceconstraint_{i}} \int_{\times_{j\in\mathbb{I}^{o}} \statespaceconstraint_{j}^{\complement}}  \nonumber\\ 
	& \qquad p(\stseq_{\settimeconstraint^{i} \cup \settimeconstraint^{o}} | \tb , \td) \delta_{\{\traj\}}(\setY) \diff \stseq_{\settimeconstraint^{i}} \diff \stseq_{\settimeconstraint^{o}} \nonumber \\
	= &  \delta_{\emptyset}(\setY) (1- \Pr(x_{\timeconstraint_{i}} \in \statespaceconstraint_{i}, \  \forall i \in \mathbb{I}_{\tb}^{\td} | \settimeconstraint_{\tb}^{\td} \neq \emptyset )) \\
	& + \Pr(x_{\timeconstraint_{i}} \in \statespaceconstraint_{i}, \  \forall i \in \mathbb{I}_{\tb}^{\td} | \settimeconstraint_{\tb}^{\td} \neq \emptyset ) \nonumber \\
	& \quad \times \int p^{\setconstraint}(\stseq_{\timeseq{\tb}{\td}}|\tb,\td)  \delta_{\{\traj\}}(\setY) \diff \stseq_{\timeseq{\tb}{\td}} \nonumber
\end{align}

\subsection{Disjunct constraints}

When the spatial constraints are disjunct, the state has to be in at least one constraint region.

\begin{align}
	& \int p(\stseq_{\timeseq{\tb}{\td}}) \delta_{\tau^{\setconstraint}(\{\traj\})}(\setY) \diff \stseq_{\timeseq{\tb}{\td}} \nonumber \\
	& = \sum_{\substack{\mathbb{I}^{i} \uplus \mathbb{I}^{o} = \mathbb{I}_{\tb}^{\td}: \\ \mathbb{I}^{o}=\emptyset}}  \int_{\times_{i\in\mathbb{I}^{i}} \statespaceconstraint_{i}} \int_{\times_{j\in\mathbb{I}^{o}} \statespaceconstraint_{j}^{\complement}}  \\ 
	& \qquad p(\stseq_{\settimeconstraint^{i} \cup \settimeconstraint^{o}} | \tb , \td) \delta_{\emptyset}(\setY) \diff \stseq_{\settimeconstraint^{i}} \diff \stseq_{\settimeconstraint^{o}} \nonumber \\
	& \quad + \sum_{\substack{\mathbb{I}^{i} \uplus \mathbb{I}^{o} = \mathbb{I}_{\tb}^{\td}: \\ \mathbb{I}^{o}\neq\emptyset}} \int_{\times_{i\in\mathbb{I}^{i}} \statespaceconstraint_{i}} \int_{\times_{j\in\mathbb{I}^{o}} \statespaceconstraint_{j}^{\complement}}  \nonumber\\ 
	& \qquad p(\stseq_{\settimeconstraint^{i} \cup \settimeconstraint^{o}} | \tb , \td) \delta_{\{\traj\}}(\setY) \diff \stseq_{\settimeconstraint^{i}} \diff \stseq_{\settimeconstraint^{o}} \nonumber \\
	= &  \delta_{\emptyset}(\setY)  \Pr(x_{\timeconstraint_{i}} \in \statespaceconstraint_{i}^{\complement}, \  \forall i \in \mathbb{I}_{\tb}^{\td} | \settimeconstraint_{\tb}^{\td} \neq \emptyset )  \\
	& +\left(1- \Pr(x_{\timeconstraint_{i}} \in \statespaceconstraint_{i}^{\complement}, \  \forall i \in \mathbb{I}_{\tb}^{\td} | \settimeconstraint_{\tb}^{\td} \neq \emptyset ) \right)  \nonumber \\
	& \times \sum_{\substack{\mathbb{I}^{i} \uplus \mathbb{I}^{o} = \mathbb{I}_{\tb}^{\td}: \\ \mathbb{I}^{i} \neq \emptyset}} w^{\mathbb{I}^{i} , \mathbb{I}^{o}}  \int p^{\mathbb{I}^{i},\mathbb{I}^{o}}(\stseq_{\timeseq{\tb}{\td}}) \delta_{\{\traj\}}(\setY) \diff \stseq_{\timeseq{\tb}{\td}}
\end{align}

\ifCLASSOPTIONcaptionsoff
  \newpage
\fi

\bibliographystyle{IEEEtran}
\bibliography{references,traj}


\end{document}